\documentclass{llncs} 
\pdfoutput=1

\usepackage[utf8]{inputenc}
\usepackage{float}
\usepackage{placeins}
\usepackage{graphicx}
\usepackage{amsmath, amssymb}

\usepackage{hyperref}

\usepackage{rotating}
\usepackage{multirow}
\usepackage{colortbl}
\usepackage{mdframed}
\usepackage{environ}
\usepackage{varwidth}

\usepackage[super]{nth}
\usepackage{pdflscape}

\usepackage[T1]{fontenc}
\usepackage{lmodern}

\usepackage{amsmath}
\usepackage{amsfonts}
\usepackage{amssymb}
\usepackage{mathpartir}
\usepackage{bm}

\newlength{\EqBoxWidthTweak}%

\NewEnviron{eqbox}[1][]{%
    \setlength{\EqBoxWidthTweak}{\dimexpr%
        +\mdflength{innerleftmargin}
        +\mdflength{innerrightmargin}
        +\mdflength{leftmargin}
        +\mdflength{rightmargin}
        }%
    \savebox0{%
        \begin{varwidth}{\dimexpr\linewidth-\EqBoxWidthTweak\relax}%
            \BODY
        \end{varwidth}%
    }%
    \begin{mdframed}[userdefinedwidth=\dimexpr\wd0+\EqBoxWidthTweak\relax, #1]
        \usebox0
    \end{mdframed}
}

\spnewtheorem{prop}{Property}{\bfseries}{\itshape}

\begin{document}

%
%

\title{Privacy Architectures: Reasoning\\About Data Minimisation and Integrity\protect\footnote{The final publication is available at link.springer.com (URL not yet available).}}

\author{Thibaud Antignac \and Daniel Le M\'etayer}

\institute{Inria, University of Lyon, France\\
\email{\{thibaud.antignac,daniel.le-metayer\}@inria.fr}}



\maketitle

%
%

\begin{abstract}
Privacy by design will become a legal obligation in the European Community if the Data Protection Regulation eventually gets adopted. However, taking into account privacy requirements in the design of a system is a challenging task. We propose an approach based on the specification of privacy architectures and focus on a key aspect of privacy, data minimisation, and its tension with integrity requirements. We illustrate our formal framework through a smart metering case study.
\end{abstract}

%
%

\section{Introduction}
\label{introduction}

The philosophy of privacy by design is that privacy should not be treated as an afterthought but as a first-class requirement in the design of IT systems. Privacy by design will  become a legal obligation in the European Community if the Data Protection Regulation~\cite{european-parliament:2014} eventually gets adopted. However, from a technical standpoint privacy by design is a challenging endeavour: first, privacy is a multi-faceted notion stemming from a variety of principles\footnote{These principles include collection limitation, data quality, purpose specification, use limitation, security, openness, individual participation, accountability, etc.} which are generally not defined very precisely; in addition, these requirements may be (or may seem to be) in tension with other requirements such as functional requirements, ease of use or performances. To implement these requirements, a wide array of privacy enhancing technologies (PETs) are available\footnote{For example homomorphic encryption, zero-knowledge proof, secure multi-party computation, private information retrieval, anonymous credentials, anonymous communication channels, etc.}. Each of these techniques provides different guarantees based on different assumptions and therefore is suitable in different contexts. As a result, it is quite complex for a software engineer to make informed choices among all these possibilities and to find the most appropriate combination of techniques to solve his own requirements. Solutions have been proposed in different application domains such as smart metering~\cite{garcia:2011,rial:2010}, pay-as-you-drive \cite{balasch:2010,jonge:2008}, or location-based systems \cite{krumm:2009} but the next challenge in this area is to go beyond individual cases and to establish sound foundations and methodologies for privacy by design~\cite{diaz:2009,tschantz:2009}. In this paper, we advocate the idea that privacy by design should be addressed at the architectural level, because it makes it possible to abstract away unnecessary details, and should be supported by a formal model. 
The fact that not all aspects of privacy are susceptible to formalisation is not a daunting obstacle to the use of formal methods for privacy by design: the key issue is to be able to build appropriate models for the aspects of privacy that are prone to formalisation and involve complex reasoning. Data minimisation, which is one of the key principles of most privacy guidelines and regulations, is precisely one of these aspects. Data minimisation stipulates that the collection and processing of personal data should always be done with respect to a particular purpose and the amount of data strictly limited to what is really necessary to achieve the purpose~\cite{european-parliament:2014}.

In this paper, data minimisation requirements are expressed as properties defining for each stakeholder the information that he is (or is not) allowed to know. Data minimisation would not be so difficult to achieve if other, sometimes conflicting, requirements did not have to be met simultaneously. Another common requirement, which  we call ``integrity'' in the sequel, is the fact that some stakeholders may require guarantees about the correctness of the result of a computation. In fact, the tension between data minimisation and integrity is one of the delicate issues to be solved in many systems involving personal data. 

In Section~\ref{sec:architectures} we propose a language to define privacy architectures. In Section~\ref{sec:logic}, we introduce a logic for reasoning about architectures and show the correctness and completeness of its axiomatisation. This axiomatisation is used in Section~\ref{sec:case-study} to prove that an example of smart metering architecture meets the expected privacy and minimisation requirements. Section~\ref{sec:related-works} discusses related work and Section~\ref{sec:directions-for-further-work} outlines directions for further research.

\section{Privacy Architectures}
\label{sec:architectures}

Many definitions of architectures have been proposed in the literature. In this paper, we adopt a definition inspired by~\cite{bass:2012}\footnote{This definition is a generalisation (to system architectures) of the definition of software architectures proposed in~\cite{bass:2012}.}: \textit{The architecture of a system is the set of structures needed to reason about the system, which comprise software and hardware elements, relations among them and properties of both.} The atomic components of an architecture are coarse-grain entities such as modules, components or connectors. 
In the context of privacy, the components are typically the PETs themselves and the purpose of the architecture is their combination to achieve the requirements of the system.

The meaning of the requirements considered here (minimisation and integrity) depends on the purpose of the data collection, which is equated to the expected functionality of the system here. In the sequel, we assume that this functionality is expressed as the computation of a set of equations\footnote{Which is typically the case for systems involving integrity requirements.} $\Omega$ such that $\Omega = \left\{ \tilde{X} = T \right\}$ with terms $T$ defined as shown in Table~\ref{tab:eq:terms}. $\tilde{X}$ represents (potentially indexed) variables and $X$ simple variables ($X \in \textit{Var}$), $k$ index variables ($k \in \textit{Index}$), $Cx$ constants ($Cx \in \textit{Const}$), $Ck$ index constants ($Ck \in \mathbb{N}\footnote{Set of natural numbers.}$), $F$ functions ($F \in Fun$) and $\odot F (X)$ is the iterative application of function $F$ to the elements of the array denoted by $X$ (e.g. sum of the elements of $X$ if $F$ is equal to $+$). We assume that each array variable $X$ represents an array of fixed size $Range(X)$.

\vspace{-0.25cm}
\begin{table}[htbp!]
\begin{mdframed}
\vspace{-0.4cm}
\begin{align*}
    T & ::= \, \tilde{X} \mid Cx  \mid F (T_1, \dots, T_n) \mid \odot F (X) \\
    \tilde{X} & ::= \, X \mid X_K \\
    K & ::= \, k \mid Ck
\end{align*}
\end{mdframed}
\caption{Term Language.}
\label{tab:eq:terms}
\end{table}
\vspace{-0.75cm}

In the following subsections, we introduce our privacy architecture language (Subsection~\ref{ssec:language}) and its semantics (Subsection~\ref{ssec:semantics}). 

\subsection{Privacy Architecture Language}
\label{ssec:language}

We define an architecture as a set of components $C_i$, $i \in [1,\ldots,n]$ associated with relations describing their capabilities. These capabilities depend on the set of available PETs. For the purpose of this paper, we consider the architecture language described in Table~\ref{tab:eq:arch}.

\vspace{-0.25cm}
\begin{table}[htbp!]
\begin{mdframed}
\vspace{-0.4cm}
\begin{align*}
     A ::= \, &    \{R\} \\
     R ::= \, &\textit{Has}_i \left( \tilde{X} \right) &&&&\hspace{-1.8cm} \mid \, \textit{Receive}_{i,j} \left( \{\textit{S}\}, \{\tilde{X}\} \right)\\ 
             \mid\, &\textit{Compute}_i \left( \tilde{X} = T \right) &&&&\hspace{-1.8cm} \mid\, \textit{Check}_i \left(\{ \textit{Eq} \} \right)\\
             \mid\, &\textit{Verif}^{\textit{Proof}}_i \left( \textit{Pro} \right) &&&&\hspace{-1.8cm} \mid\, \textit{Verif}^{\textit{Attest}}_i \left( \textit{Att} \right)\\
             \mid\, &\textit{Spotcheck}_{i,j} \left( X_k, \textit{Eq} \right) &&&&\hspace{-1.8cm} \mid\, \textit{Trust}_{i,j}\\
             \\
    \textit{S} ::= \, &\textit{Pro} \mid \textit{Att} &&&\hspace{-1.8cm} \textit{Att} ::= \, &\textit{Attest}_i \left( \{\textit{Eq}\} \right)&&\\
    \textit{Pro} ::= \, &\textit{Proof}_i \left( \left\{ \textit{P} \right\} \right) &&&\hspace{-1.8cm} \textit{Eq} ::= \, &\textit{T}_1\; \textit{Rel}\; \textit{T}_2&&\\
    \textit{P} ::= \, &\textit{Att} \mid \textit{Eq} &&&\hspace{-1.8cm} \textit{Rel} ::= \, &= \,\mid\, < \,\mid\, > \,\mid\, \leq \,\mid\, \geq &&
\end{align*}
\end{mdframed}
\caption{Privacy Architecture Language.}
\label{tab:eq:arch}
\end{table}
\vspace{-0.75cm}

\begin{figure}[htbp!]
    \begin{center}
        \includegraphics[width=1\textwidth]{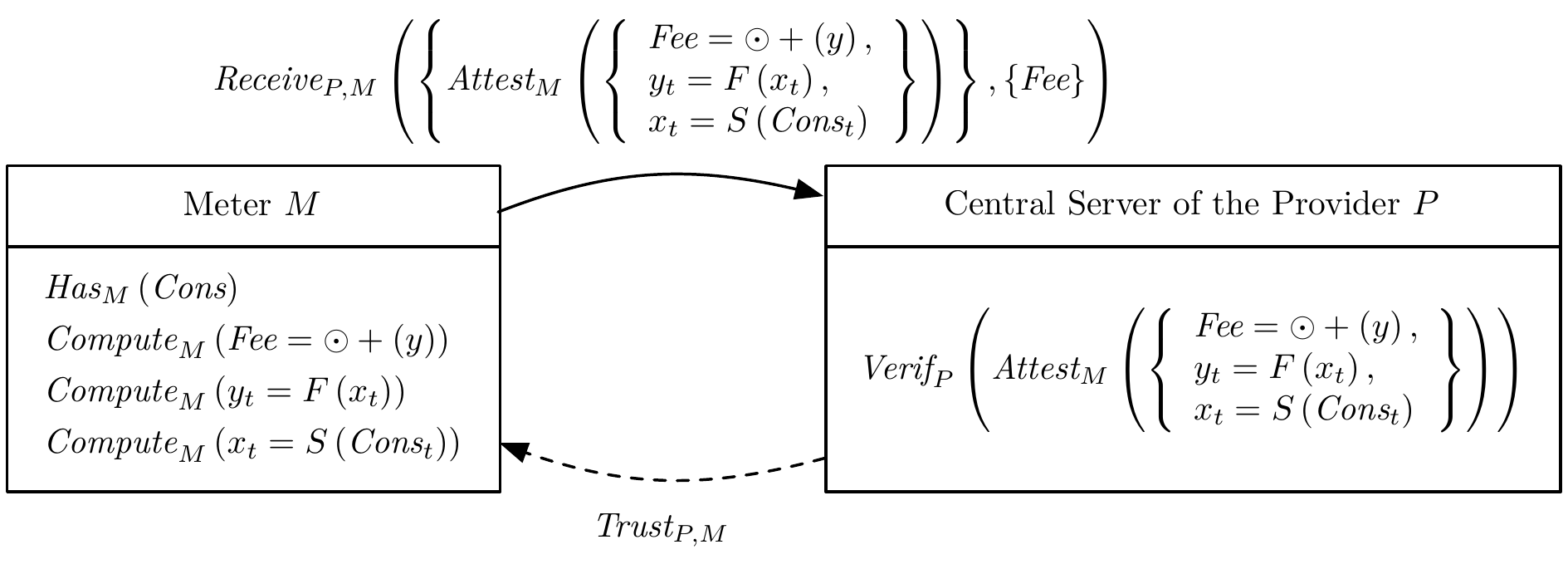}
    \end{center}
        \vspace{-0.5cm}
    \caption{Example of smart metering architecture.}
    \label{fig:trust-blind}
\end{figure}

Subscripts $i$ and $j$ are component indexes and the notation $\{Z\}$ is used to define a set of terms of category $Z$.  $\textit{Has}_i(\tilde{X})$ expresses the fact that variable $\tilde{X}$ is an input variable located at component $C_i$ (e.g. sensor or meter) and $\textit{Receive}_{i,j}(\{S\}, \{\tilde{X}\})$ specifies that component $C_i$ can receive from component $C_j$ messages consisting of a set of statements $\{S\}$ and a set of variables $\{\tilde{X}\}$. A statement can be either a proof of a set of properties $P$ (denoted by $\textit{Proof}_i \left( \left\{ P \right\} \right)$) or an attestation (denoted by $\textit{Attest}_i(\{Eq\})$), that is to say a simple declaration by a component $C_i$ that properties $Eq$ are true. A component can also compute a variable defined by an equation $\tilde{X} = T$ (denoted by $\textit{Compute}_i(\tilde{X} = T)$), check that a set of properties $\textit{Eq}$ holds (denoted by $\textit{Check}_i\left(\{\textit{Eq}\}\right)$), verify a proof of a property $\textit{Pro}$ received from another component (denoted by $\textit{Verif}^{Proof}_i (\textit{Pro})$), verify the origin of an attestation (denoted by $\textit{Verif}^{\textit{Attest}}_i (\textit{Att})$), or perform a spotcheck. A spotcheck, which is denoted by $\textit{Spotcheck}_{i,j} (X_k, \textit{Eq})$, is the request from a component $C_j$ of a value $X_k$ taken from array $X$ and the verification that this value satisfies property $\textit{Eq}$. Primitive properties $\textit{Eq}$ are simple equations on terms $T$. Last but not least, trust assumptions  are expressed using ${\textit{Trust}}_{i,j}$ (meaning that component $C_i$ trusts component $C_j$). In the sequel, we use $\Gamma$ to denote the set of architectures following the syntax of Table~\ref{tab:eq:arch}. Architectures can also be defined using graphical representations. As an illustration, Figure~\ref{fig:trust-blind} displays a simple architecture involving a meter $M$ and the central server of a provider $P$. The meter plays both the role of a sensor providing the input consumption values ($\textit{Has}_M(\textit{Cons}_t)$) and the role of a secure element computing the fee. Because the provider trusts the meter ($\textit{Trust}_{P,M}$), it merely checks the certificate $\textit{Attest}_M \left( \left\{ \textit{Fee} = \odot + \left( y \right), y_t = F \left( x_t \right), x_t = S \left( \textit{Cons}_t \right) \right\} \right)$ sent by the meter. 

\vspace{-0.25cm}
    \vspace{-0.25cm}

Strictly speaking, we should introduce a notion of actor and a relationship between actors and the components that are under their control but, for the sake of brevity (and without loss of generality\footnote{The fact that an actor controls several components can be expressed through a trust relationship.}), we do not distinguish between components and actors here.

Architectures provide an abstract, high-level view of a system: for example, we do not express at this level the particular method (for example, a zero-knowledge proof protocol) used by a component to build a proof ($\textit{Proof}_i \left( \left\{ P \right\} \right)$) or to verify it, or to check that another component has actually certified (attested) a property ($\textit{Verif}^{\textit{Attest}}_i (\textit{Att})$). Another main departure from protocol specification languages is that we do not have any specific ordering or notion of sequentiality here, even though functional dependencies introduce implicit constraints in the events of the system, as discussed below. The objective is to express and reason about the main design choices rather than to cover all the development steps.

\subsection{Privacy Architectures Semantics}
\label{ssec:semantics}

The definition of the semantics of an architecture is based on its  set of compatible traces. A trace is a sequence of high-level events occurring in the system as presented in Table~\ref{tab:eq:events}. Events can be seen as instantiated relations of the architecture. For example, a $\textit{Receive}_{i,j} \left( \{\textit{S}\}, \{\tilde{X}:V\} \right)$ event specifies the values $V$ of the variables $\tilde{X}$ received by $C_i$. Similarly, $ \textit{Spotcheck}_{i,j} \left( X_{\textit{Ck}}:V, \{\textit{Eq}\} \right)$ specifies the specific index $\textit{Ck}$ (member of $\mathbb{N}$) chosen by $C_i$ for the spotcheck and the value $V$ of $X_{\textit{Ck}}$. All variable indexes occurring in events, except for variables occurring in the properties of $\textit{Receive}_{i,j}$, $\textit{Verif}^{\textit{Proof}}_{i}$, $\textit{Verif}^{\textit{Attest}}_{i}$, and $\textit{Spotcheck}_{i,j}$, must belong to $\mathbb{N}$.

\vspace{-0.25cm}
\begin{table}[htbp!]
\begin{mdframed}
\vspace{-0.4cm}
\begin{align*}
\theta ::= \, & \textit{Seq} (\epsilon) \\ 
 \epsilon ::= \, &\textit{Has}_i \left( \tilde{X}:V \right) &&\hspace{-2.8cm} \mid \, \textit{Receive}_{i,j} \left( \{\textit{S}\}, \{\tilde{X}:V\} \right)\\ 
             \mid\, &\textit{Compute}_i \left( \tilde{X} = T \right) &&\hspace{-2.8cm} \mid\, \textit{Check}_i \left( \{ \textit{Eq} \} \right)\\
             \mid\, &\textit{Verif}^{\textit{Proof}}_i \left( \textit{Pro} \right) &&\hspace{-2.8cm} \mid\, \textit{Verif}^{\textit{Attest}}_i \left( \textit{Att} \right)\\
             \mid\, &\textit{Spotcheck}_{i,j} \left( X_{\textit{Ck}}:V, \{\textit{Eq}\} \right)   
\end{align*}
\end{mdframed}
\caption{Events and traces.}
\label{tab:eq:events}
\end{table}
\vspace{-0.75cm}

In the following, we consider only consistent architectures and consistent traces. An architecture is said to be consistent if each variable can be computed (or can be initially possessed, as expressed by $\textit{Has}_i$) by a single component, a component cannot receive a variable from different sources, a component computing a variable or checking a property can receive or compute all the necessary input variables (variables occuring in $T$ for $Compute_i (\tilde{X} = T)$, in $\textit{Eq}$ for $\textit{Check}_i(\textit{Eq})$), a component can only verify properties that it can receive from another component, etc. The same kind of consistency assumptions apply to traces, in addition to ordering consistency properties (variables and properties are not used before being received or computed).

We use $\textit{Event}$ to denote the set of events $\epsilon$ and $\textit{Trace}$ to denote the set of consistent traces $\theta$.

\begin{definition}[Compatibility] A trace $\theta$ of length $\overline{\theta}$  is  compatible with an architecture $A$ if and only if:

\begin{align*}
\forall a \in [1,\overline{\theta}], \, & \text{ if }  \theta_a \neq \textit{Compute}_i \left( \tilde{X} = T \right) \text{ then } \exists \alpha \in\, A, \mathcal{C} (\theta_a, \alpha)  \text{ and} \\
& \text{ if } \theta_a = \textit{Spotcheck}_{i,j} \left( X_{\textit{Ck}}:V, \{\textit{Eq}\} \right) \\ 
& \hphantom{\text{ space }} \text{ then } \forall b \in [1,\overline{\theta}], b \neq a \Rightarrow \forall k^{\prime}, V^{\prime}, Eq^{\prime}, \, \\
& \hphantom{\text{ spacespacespacespacespacespac }} \theta_b \neq \textit{Spotcheck}_{i,j} \left( X_{\textit{k}^{\prime}}:V^{\prime}, \{\textit{Eq}^{\prime}\} \right)
\end{align*}
where $\mathcal{C} (\epsilon, \alpha)$ holds if and only if $\epsilon$ can be obtained from $\alpha$ by adding specific values $V$ for variables and instantiating index variables to integer values. 
\end{definition}

The first condition in the definition of compatibility states that, except for compute events, only events which are instantiations of  components of the architecture $A$ can appear in the trace $\theta$. 
The rationale for excepting compute events is the need to express the potential actions of a curious agent trying to derive the value of a variable $\tilde{X}$ from the values of variables that he already has. As a result, compatible traces may include computations that are not contemplated by the architecture, provided that the component possesses all the variables necessary to perform this computation (consistency assumption). The adversary model considered here includes computation of new variables, erroneous computations, and communication of incorrect values, which corresponds to Dolev-Yao attacks for internal stakeholders (except they cannot break the protocol). The second condition expresses the fact that spotchecks can be performed only once. This condition could be relaxed through the introduction of an additional threshold parameter $t$ to express the fact that up to $t$ spotchecks are possible. We denote by $T(A)$ the set of compatible traces of an architecture $A$.

In order to define the semantics of events, we introduce first the notion of state of a component:
\begin{align*}
\textit{State} = \, &\left( \textit{State}_V \, \times \, \textit{State}_P \times \, \textit{State}_P \right) \cup \left\{ \textit{Error} \right\}\\
\textit{State}_V = \, &\left( \textit{Var} \rightarrow \textit{Val}_{\bot} \right)\\
\textit{State}_{P} = \, &\left\{ \{Eq\} \cup \{\textit{Trust}_{i,j}\} \right\}
\end{align*}

The state of a component is either the error state $\textit{Error}$ or a triple made of a variable state assigning a value (or the undefined value $\bot$\footnote{Please note that $\bot$ is used to denote undefined values, that is to say values which have not been set, as opposed to error values (e.g. division by zero or type errors). We do not consider computation error values here.}) to each variable and two property states: the first one defines the set of properties known by the component and the second one the set of properties believed by the component (after a spotcheck). 
In the sequel, we use $\sigma$ to denote the global state (state of the components $\langle C_1, \dots, C_n \rangle$) defined on $\textit{State}^n$. 
The initial state for an architecture $A$ is denoted by $\textit{Init}^A = \langle \textit{Init}^A_1, \ldots , \textit{Init}^A_{n} \rangle$ with:
\[  \forall i \in [1,n], Init^A_i = (\textit{Empty}, \{\textit{Trust}_{i,j} | \textit{Trust}_{i,j}  \in A \}, \emptyset) \]
where $\textit{Empty}$ denotes the empty variable state ($\forall X \in \textit{Var}, \textit{Empty}(X) = \bot$). The only information contained in the initial state is the trust properties specified by the architecture.

The semantics function $S_T$ is defined in Table~\ref{tab:eq:semantics-of-traces-of-events}. It specifies the impact of a trace on the state of each component $C_i$. It is defined as an iteration through the trace with function $S_E$ defining the impact of each type of event on the states of the components.

\begin{table}[htb!]
\begin{mdframed}
\begin{align*}
S_T : \,  \textit{Trace} \,  \times \,  \textit{State}^n & \rightarrow   \textit{State}^n \\
S_E : \, \textit{Event} \,  \times \,  \textit{State}^n & \rightarrow  \textit{State}^n \\  
S_T \left(\langle\rangle, \sigma\right) & =   \sigma \\ 
S_T \left(\epsilon.\theta, \sigma\right) &  =  S_T(\theta,S_E(\epsilon,\sigma)) \\
S_E \left(\textit{Has}_i \left( \tilde{X}:V \right), \sigma\right) & = \sigma [\sigma_i / (\sigma^v_i[\tilde{X}/V], \sigma^{pk}_i, \sigma^{pb}_i) ] \\ 
S_E \left(\textit{Receive}_{i,j} \left( \{\textit{S}\}, \{\tilde{X}:V\} \right), \sigma\right) & =   \sigma [\sigma_i / (\sigma^v_i[\{\tilde{X}/V\}], \sigma^{pk}_i, \sigma^{pb}_i) ] \\
S_E \left(\textit{Compute}_i \left( \tilde{X}= T \right), \sigma\right) & =  \sigma [\sigma_i / (\sigma^v_i[\tilde{X}/\varepsilon(T,\sigma^v_i)], \sigma^{pk}_i \cup \{\tilde{X} = T\}, \sigma^{pb}_i) ]\\ 
S_E \left(\textit{Check}_i \left( E \right), \sigma\right) & = \sigma [\sigma_i / (\sigma^v_i, \sigma^{pk}_i \cup E, \sigma^{pb}_i) ] \\
& \hphantom{\text{ space }} \text{ if } \forall Eq \in E, \varepsilon(Eq,\sigma^v_i) = \textit{True} \\
&  = \sigma[\sigma_i/\textit{Error}] \text{ otherwise} \\
S_E \left(\textit{Verif}^{\textit{Proof}}_i \left( \textit{Proof}_j(E) \right), \sigma\right) & =  \sigma [\sigma_i / (\sigma^v_i, \sigma^{pk}_i \cup \{Eq |  Eq \in E \text{ or } \\
& \hphantom{\text{ spacespacespacesp }} ( \textit{Attest}_{j^{\prime}}(E^{\prime}) \in E \text{ and } \\
& \hphantom{\text{ spacespacespacesp }} Eq \in E^{\prime} \text{ and } \\
& \hphantom{\text{ spacespacespacesp }} \textit{Trust}_{i,j^{\prime}} \in  \sigma^{pk}_i ) \}, \sigma^{pb}_i)] \\
& \hphantom{\text{ space }} \text{ if } \overline{\textit{Verif}}^{\textit{Proof}}(\left( E \right), \sigma^v_i) = \textit{True} \\
&  = \sigma[\sigma_i/\textit{Error}] \text{ otherwise} \\
S_E \left(\textit{Verif}^{\textit{Attest}}_i \left( \textit{Attest}_j(E) \right), \sigma\right) & = \sigma [\sigma_i / (\sigma^v_i, \sigma^{pk}_i \cup \{Eq |  Eq \in E \text{ and } \\
& \hphantom{\text{ spacespacespacespace }}\textit{Trust}_{i,j} \in  \sigma^{pk}_i\}, \sigma^{pb}_i)] \\
& \hphantom{\text{ space }} \text{ if } \overline{\textit{Verif}}^{\textit{Attest}}(\left( E \right), \sigma^v_i) = \textit{True} \\
&  = \sigma[\sigma_i/\textit{Error}] \text{ otherwise} \\
S_E \left(\textit{Spotcheck}_{i,j} \left( X_{\textit{Ck}}:V, E \right), \sigma\right) & =  \sigma [\sigma_i / (\sigma^v_i[X_{\textit{Ck}}/V], \sigma^{pk}_i, \sigma^{pb}_i \cup E)] \\
& \hphantom{\text{ space }} \text{ if } \forall Eq \in E, \\
& \hphantom{\text{ spacespa }} \varepsilon(Eq[k/Ck],\sigma^v_i[X_{\textit{Ck}}/V]) = \textit{True} \\
&  = \sigma[\sigma_i/\textit{Error}] \text{ otherwise} 
\end{align*}
\end{mdframed}
\caption{Semantics of traces of events.}
\label{tab:eq:semantics-of-traces-of-events}
\end{table}

The notation $\epsilon.\theta$ is used to denote a trace whose first element is $\epsilon$ and the rest of the trace is $\theta$. Each event modifies only the state of the component $C_i$. This modification is expressed as $\sigma [\sigma_i / (v, \textit{pk}, \textit{pb}) ]$ (or $\sigma [\sigma_i / \textit{Error} ]$ in the case of the error state) that replaces the variable and property components of the state of $C_i$ by $v$, $\textit{pk}$, and $\textit{pb}$ respectively. We assume that no event $\theta_{a^{\prime}}$ with $a^{\prime} > a$ involves component $C_i$ if its state $\sigma_i$ is equal to $\textit{Error}$ after the occurrence of $\theta_{a}$ (in other words, any error in the execution of a component causes this component to stop).

The effect of $\textit{Has}_i$ and $\textit{Receive}_{i,j}$ on the variable state of component $C_i$ is the replacement of the values of the variables $\tilde{X}$ by new values $V \in \textit{Val}$, which is denoted by $\sigma^v_i[\tilde{X}/V]$.

The effect of ${\textit{Compute}}_i(\tilde{X}=T)$ is to set the variable $\tilde{X}$ to the evaluation of the value of $T$ in the current variable state $\sigma^v_i$ of $C_i$, which is defined by $\varepsilon(T,\sigma^v_i)$. ${\textit{Spotcheck}}_i(X_{\textit{Ck}}:V, E)$ sets the value of $X_{\textit{Ck}}$ to $V$. The other events do not have any effect on the variable state of $C_i$. The value of a variable replaced after the occurrence of an event must be $\bot$ before its occurrence\footnote{Because we consider only consistent traces. A value different from $\bot$ would mean that the variable is computed or set more than once.}. We assume that it is different from $\bot$ and does not involve any $\bot$) after the event\footnote{In other words, input values and results of computations are fully defined.}.

Most events also have an effect on the property states. This effect is the addition to the property states of the new knowledge or belief provided by the event. For ${\textit{Compute}}_i \left( \tilde{X} = T \right)$, this new knowledge is the equality $\tilde{X}=T$; for the $\textit{Check}_i$, $\textit{Verif}_i$, and $\textit{Spotcheck}_{i,j}$ events, the new knowledge is the properties checked or verified. In all cases except for $\textit{Spotcheck}_{i,j}$ these properties are added to the $\textit{pk}$ property state because they are known to be true by component $C_i$; in the case of $\textit{Spotcheck}_{i,j}$ the properties are added to the $\textit{pb}$ property state because they are believed by $C_i$: they have been checked on a sample value $X_{\textit{Ck}}$ but might still be false for some other $X_{\textit{k}}$. The only guarantee provided to $C_i$ by $\textit{Spotcheck}_{i,j}$ is that $C_i$ has always the possibility to detect an error (but he has to choose an appropriate index, that is to say an index that will reveal the error).  

Functions $\overline{\textit{Verif}}^{\textit{Proof}}$ and $\overline{\textit{Verif}}^{\textit{Attest}}$ define the semantics of the corresponding verification operations. As discussed above, we do not enter into the internals of the proof and attestation verifications here and just assume that only true properties are accepted by $\overline{\textit{Verif}}^{\textit{Proof}}$ and only attestations provided by the authentic sender are accepted by $\overline{\textit{Verif}}^{\textit{Attest}}$.
The distinctive feature of ${\textit{Verif}}^{\textit{Attest}}_i$ events is that they generate new knowledge only if the author of the attestation can be trusted (hence the $\textit{Trust}_{i,j} \in A$ condition). 

Let us note also that $\textit{Receive}_{i,j}$ events do not add any new knowledge by themselves because the received properties have to be verified before they can be added to  the property states.

We can now define the semantics of an architecture $A$ as the set of the possible states produced by compatible traces.

\begin{definition}[Semantics of architectures.]
The semantics of an architecture $A$ is defined as:
$ \mathcal{S}(A) = \{   \sigma \in \textit{State}^n \, | \, \exists \theta \in T(A), S_T(\theta, \textit{Init}^A) = \sigma \}   $.
\end{definition}

In the following, we use $ \mathcal{S}_i(A)$ to denote the subset of $ \mathcal{S}(A)$ containing only states which are well defined for component $C_i$:
 $\mathcal{S}_i(A) = \{   \sigma \in \mathcal{S}(A) \, | \,   \sigma_i  \neq  \textit{Error} \} $. The prefix ordering on traces gives rise to the following ordering on states:  $\forall \sigma \in \mathcal{S}_i(A), \forall \sigma^{\prime} \in \mathcal{S}_i(A), \sigma \geq_i \sigma^{\prime} \Leftrightarrow \exists \theta \in T(A), \exists \theta^{\prime} \in T(A), \sigma = S_T(\theta, \textit{Init}^A), \sigma^{\prime} = S_T(\theta^{\prime}, \textit{Init}^A), \text{ and } \theta^{\prime} \text{ is a prefix of } \theta$.

\section{Privacy Logic}
\label{sec:logic}

Because privacy is closely connected with the notion of knowledge, epistemic logics form an ideal basis to reason about privacy properties. Epistemic logics~\cite{fagin:2004} are a family of modal logics using a knowledge modality usually denoted by $K_i\left(\psi\right)$ to denote the fact that agent $i$ knows the property $\psi$. However standard epistemic logics based on possible worlds semantics suffer from a weakness which makes them unsuitable in the context of privacy: this problem is often referred to as ``logical omniscience''~\cite{halpern:2007}. It stems from the fact that agents know all the logical consequences of their knowledge (because these consequences hold in all possible worlds). An undesirable outcome of logical omniscience would be that, for example, an agent knowing the hash $H\left(v\right)$ of a value $v$ would also know $v$. This is obviously not the intent in a formal model of privacy where hashes are precisely used to hide the original values to the recipients. This issue is related to the fact that standard epistemic logics do not account for limitations of computational power.

Therefore it is necessary to define dedicated epistemic logics to deal with different aspects of privacy and to model the variety of notions at hand (e.g. knowledge, zero-knowledge proof, trust, etc.). 
In this paper, we follow the ``deductive algorithmic knowledge'' approach~\cite{fagin:2004,pucella:2004} in which the explicit knowledge of a component $C_i$ is defined as the knowledge that this component can actually compute using his own deductive system $\triangleright_i$. The deductive relation $\triangleright_i$ is defined here as a relation between a set of $\textit{Eq}$ and $\textit{Eq}$ properties: \mbox{$\left\{ \textit{Eq}_1, \dots, \textit{Eq}_n \right\} \triangleright_i \textit{Eq}_0$}. Typically, $\triangleright_i$ can be used to capture properties of the functions of the specification. For example $\left\{ h_1 = \textit{H} \left( x_1 \right), h_2 = \textit{H} \left( x_2 \right), h_1 = h_2 \right\} \triangleright_i (x_1 = x_2)$ expresses the injectivity property of a hash function $H$. Another relation, $\textit{Dep}_i$, is introduced to express that a variable can be derived from other variables. $\textit{Dep}_i \left( \tilde{X}, \left\{ \tilde{X}^1, \dots \tilde{X}^n \right\} \right)$ means that a value for $\tilde{X}$ can be obtained by $C_i$ (\mbox{$\exists F, \tilde{X} = F ( \tilde{X}^1, \dots, \tilde{X}^n )$}). The absence of a relation such as $\textit{Dep}_i \left( x_k, \left\{ y_k \right\} \right)$ prevents component $C_i$ from deriving the value of $x_k$ from the value of $y_k$, capturing the hiding property of the hash application $y_k = H \left( x_k \right)$. 

\vspace{-0.25cm}
\begin{table}[htbp!]
\begin{mdframed}
\vspace{-0.4cm}
\begin{align*}
    \phi ::= \, &\textit{Has}_i^{\textit{all}} \left( \tilde{X} \right) &&\hspace{-3.4cm} | \, \textit{Has}_i^{\textit{none}} \left( \tilde{X} \right) &&\hspace{-3.3cm} | \, \textit{Has}_i^{\textit{one}} \left( \tilde{X} \right)\\
        | \, &\textit{K}_i \left( \textit{Eq} \right) &&\hspace{-3.4cm} | \, \textit{B}_i \left( \textit{Eq} \right) &&\hspace{-3.3cm} | \, \phi_1 \wedge \phi_2\\
    \textit{Eq} ::= \, &\textit{T}_1\; \textit{Rel}\; \textit{T}_2 \mid Eq_1 \wedge Eq_2
\end{align*}
\end{mdframed}
\caption{Architecture logic.}
\label{tab:eq:logic}
\end{table}
\vspace{-0.75cm}

This logic involves two modalities, denoted by $K_i$ and $B_i$, which represent respectively  knowledge and  belief properties of a component $C_i$. Please note that the $Eq$ notation (already used in the language of architectures) is overloaded, without ambiguity: it is used to denote conjunctions (rather than sets) of primitive relations in the logic. 
The logic can be used  to express useful properties of architectures: for example $\textit{Has}_i^{\textit{all}}(\tilde{X})$ expresses the fact that component $C_i$ can obtain or derive (using its deductive system $\triangleright_i$) the value of $\tilde{X}_k$ for all $k$ in $\textit{Range}(X)$. $\textit{Has}_i^{\textit{one}}(\tilde{X})$ expresses the fact that component $C_i$ can obtain or derive the value of $X_k$ for at most one $k$ in $\textit{Range}(X)$. Finally, $\textit{Has}_i^{\textit{none}}(\tilde{X})$ is the privacy property stating that $C_i$ does not know any $X_k$ value.
It should be noted that $\textit{Has}_i$ properties only inform on the fact that $C_i$ can get or derive some values for the variables but they do not bring any guarantee about the correctness of these values. 
Such guarantees can only be ensured through integrity requirements, expressed using the $K_i(\textit{Eq})$ and $B_i(\textit{Eq})$ properties. $K_i(\textit{Eq})$ means that component $C_i$ can establish the truthfulness of $\textit{Eq}$ while $B_i(\textit{Eq})$  expresses the fact that $C_i$ may test this truthfulness, and therefore detect its falsehood or believe that the property is true otherwise.


We can now define the semantics of a property $\phi$.

\begin{definition}[Semantics of properties] 
The semantics $S(\phi)$ of a property $\phi$ is defined in Table~\ref{tab:eq:semantics-of-architecture} as the set of architectures meeting $\phi$.
\end{definition}

\vspace{-0.25cm}
\begin{table}[htbp!]
\begin{mdframed}
\vspace{-0.4cm}
\begin{align*}
    A \in S\left(\textit{Has}_i^\textit{all} \left( \tilde{X} \right)\right) \, \Leftrightarrow \, & \exists \sigma \in \mathcal{S}(A), \sigma_i^v(\tilde{X}) \text{ does not contain any } \bot \\
  A \in S\left(\textit{Has}_i^\textit{none} \left( \tilde{X} \right)\right) \, \Leftrightarrow \, & \forall \sigma \in \mathcal{S}(A), \sigma_i^v(\tilde{X}) = \bot  \\
  A \in S\left(\textit{Has}_i^\textit{one} \left( \tilde{X} \right)\right) \, \Leftrightarrow \, & \forall \sigma \in \mathcal{S}(A), \sigma_i^v(\tilde{X}) = \bot \vee
 ( \sigma_i^v(\tilde{X}) = \, <v_1, \dots, v_k > \wedge \\
& \hphantom{\text{ spacespacespacespacespa }} \nexists (u, u^{\prime}),  u \neq u^{\prime}  \wedge  v_u \neq \bot  \,\wedge \\
& \hphantom{\text{ spacespacespacespacespacespacespacespaci }}  v_{u^{\prime}} \neq \bot  ) \\
  A \in S\left(\textit{K}_i \left( \textit{Eq} \right)\right) \, \Leftrightarrow \, & \forall \sigma^{\prime} \in \mathcal{S}_i(A), \exists \sigma \in \mathcal{S}_i(A), \exists \textit{Eq}^{\prime}, (\sigma \geq_i \sigma^{\prime}) \wedge ( \sigma_i^\textit{pk} \triangleright_i \textit{Eq}^{\prime}) \,\wedge \\
 & \hphantom{\text{ spacespacespacespacespacespacespacespa }} (\textit{Eq}^{\prime} \Rightarrow \textit{Eq}) \\ 
  A \in S\left(\textit{B}_i \left( \textit{Eq} \right)\right) \, \Leftrightarrow \, & \forall \sigma^{\prime} \in \mathcal{S}_i(A), \exists \sigma \in \mathcal{S}_i(A), \exists \textit{Eq}^{\prime}_1, \exists \textit{Eq}^{\prime}_2, (\sigma \geq_i \sigma^{\prime}) \wedge  \\
 & \hphantom{\text{ spacespacespacespacespace }}  (\sigma_i^{pb} \triangleright_i \textit{Eq}_1^{\prime})  \wedge (\sigma_i^{pk} \triangleright_i \textit{Eq}_2^{\prime}) \,\wedge \\
 & \hphantom{\text{ spacespacespacespacespace }} \left( \left( \textit{Eq}_1^{\prime} \wedge \textit{Eq}_2^{\prime} \right) \Rightarrow \textit{Eq} \right) \\
A \in S\left(\phi_1 \wedge \phi_2\right) \,  \Leftrightarrow \,  &  A \in S(\phi_1) \wedge A \in S(\phi_2)
\end{align*}
\end{mdframed}
\caption{Semantics of properties.}
\label{tab:eq:semantics-of-architecture}
\end{table}
\vspace{-0.75cm}

An architecture satisfies the $\textit{Has}_i^{\textit{all}}(\tilde{X})$ property if and only if $C_i$ may obtain the full value of $\tilde{X}$ in at least one compatible execution trace whereas $ \textit{Has}_i^{\textit{none}}(\tilde{X})$ holds if and only if no execution trace can lead to a state in which $C_i$ gets a value of $\tilde{X}$ (or of any part of its content if $\tilde{X}$ is an array variable). $\textit{Has}_i^{\textit{one}}(\tilde{X})$ is true if and only if no execution trace can lead to a state in which $C_i$ knows more than one of the values of the array $\tilde{X}$. The validity of $K_i(\textit{Eq})$ and $B_i(\textit{Eq})$ properties is defined with respect to correct execution traces (with respect to $C_i$) since an incorrect trace leads to a state in which an error has been detected by the component\footnote{This is a usual implicit assumption in protocol verification.}. The condition $\sigma \geq \sigma^{\prime}$ is used to discard states corresponding to incomplete traces in which the property $\textit{Eq}$ has not yet been established. As discussed above, the capacity for a component $C_i$ to derive new knowledge or beliefs is defined by its deductive system $\triangleright_i$.

In order to reason about architectures and the knowledge of the components, we introduce in Table~\ref{tab:eq:axioms} an axiomatisation of the logic presented in the previous section. The fact that an architecture $A$ satisfies a property $\phi$ is denoted by \mbox{$A \vdash \phi$}. Axioms (H1-8) and (HNO) are related to properties $\textit{Has}_i$ while axioms (K1-5) and (K$\wedge$) are related to the knowledge of the components. Axioms (B), (KB), and (B$\wedge$) handle the belief case. Finally, the remaining axioms are structural axioms dealing with the conjunctive operator.

\begin{table}[htb!]
\begin{mdframed}
\begin{align*}
 & \mathbf{H1} \inferrule{\textit{Has}_i \left( \tilde{X} \right) {\in} \, A}{A \vdash \textit{Has}_i^\textit{all} \left( \tilde{X} \right)} \hspace{0.75cm}
 \mathbf{H2} \mprset{sep=1.5em}\inferrule{\textit{Receive}_{i,j} \left( \textit{S}, E \right) {\in} \, A \\ \tilde{X} \in \{E\}}{A \vdash  \textit{Has}_i^\textit{all} \left( \tilde{X} \right)} \\
 & \mathbf{H3} \inferrule{\textit{Compute}_i \left( \tilde{X} = T \right) {\in} \, A}{A \vdash \textit{Has}_i^\textit{all} \left( \tilde{X} \right)} \hspace{0.75cm}
 \mathbf{H4} \inferrule{\textit{Spotcheck}_{i,j} \left( X_k, \textit{E} \right) \in A}{A \vdash \textit{Has}_i^\textit{one} \left( X \right)} \\
 & \mathbf{H5} \mprset{sep=1.5em}\inferrule{\textit{Dep}_i \left( \tilde{X}, \left\{ \tilde{X}^1, \dots \tilde{X}^n \right\} \right) \\ \text{for all } l \in [1,n], A \vdash \textit{Has}_i^\textit{all} \left( \tilde{X}^l \right)}{A \vdash \textit{Has}_i^\textit{all} \left( \tilde{X} \right)} \\
 & \mathbf{H6} \inferrule{\text{None of the pre-conditions of H1, H2, H3, H4, or H5 holds for } X \text{ or any } X_k}{A \vdash  \textit{Has}_i^\textit{none} \left( \tilde{X} \right)} \\
 & \mathbf{H7} \inferrule{A \vdash \textit{Has}_i^\textit{all} \left( \tilde{X} \right)}{A \vdash \textit{Has}_i^\textit{all} \left( X_k \right)} \text{ for all } k \in \textit{Range}(X) \hspace{0.75cm}
 \mathbf{HNO} \inferrule{A \vdash \textit{Has}_i^\textit{none} \left( \tilde{X} \right)}{A \vdash \textit{Has}_i^\textit{one} \left( \tilde{X} \right)} \\
 & \mathbf{H8} \inferrule{A \vdash \textit{Has}_i^\textit{none} \left( \tilde{X} \right)}{A \vdash \textit{Has}_i^\textit{none} \left( X_k \right)} \text{ for all } k \in \textit{Range}(X) \hspace{0.75cm}
 \mathbf{K1} \inferrule{\textit{Compute}_i \left( \tilde{X} = T \right) \in  A}{A \vdash K_i(\tilde{X} = T)} \\
 & \mathbf{K3} \mprset{sep=1.5em}\inferrule{\textit{Verif}^{\textit{Proof}}_i \left( \textit{Proof}_j(\textit{E}) \right) \in  A \\ \textit{Eq} \in \textit{E}}{A \vdash K_i( \textit{Eq} )} \hspace{0.75cm}
 \mathbf{K2} \mprset{sep=1.5em}\inferrule{\textit{Check}_i \left( \textit{E} \right) \in  A \\ \textit{Eq} \in \textit{E}}{A \vdash K_i(\textit{Eq} )} \\
 & \mathbf{K4} \mprset{sep=1.5em}\inferrule{\textit{Verif}^{\textit{Proof}}_i \left( \textit{Proof}_j( \textit{E}) \right) \in  A \\ \textit{Attest}_k( \textit{E}^{\prime}) \in  E \\ \textit{Trust}_{i,k} \in A \\ \textit{Eq} \in \textit{E}^{\prime}}{A \vdash K_i( \textit{Eq} )} \\
 & \mathbf{K5} \mprset{sep=1.5em}\inferrule{\textit{Verif}^{\textit{Attest}}_i \left( \textit{Attest}_j( \textit{E} ) \right) \in  A \\ \textit{Trust}_{i,j} \in A\\ \textit{Eq} \in \textit{E}}{A \vdash K_i( \textit{Eq} )} \hspace{0.75cm} 
 \mathbf{KB} \inferrule{A \vdash K_i(\textit{Eq})}{A \vdash B_i(\textit{Eq})} \\
 & \mathbf{K}\hspace{-0.2em}\wedge \mprset{sep=1.5em}\inferrule{A \vdash K_i(\textit{Eq}_1) \\ A \vdash K_i(\textit{Eq}_2)}{A \vdash K_i(\textit{Eq}_1 \wedge \textit{Eq}_2)} \hspace{0.75cm}
 \mathbf{B} \inferrule{\textit{Spotcheck}_{i,j} \left( X_k, \textit{E} \right) \in A \\ \textit{Eq} \in \textit{E}}{A \vdash B_i(\textit{Eq})} \\
 & \mathbf{K}\hspace{-0.2em}\triangleright \mprset{sep=1.5em}\inferrule{E \triangleright_i \textit{Eq}_0 \\ \text{for all } \textit{Eq} \in E, A \vdash K_i(\textit{Eq})}{A \vdash K_i(\textit{Eq}_0)} \hspace{0.75cm} 
 \mathbf{I}\hspace{-0.2em}\wedge \mprset{sep=1.5em}\inferrule{A \vdash \phi_1 \\ A \vdash \phi_2}{A \vdash \phi_1 \wedge \phi_2} \\
 & \mathbf{B}\hspace{-0.2em}\wedge \mprset{sep=1.5em}\inferrule{A \vdash B_i(\textit{Eq}_1) \\ A \vdash B_i(\textit{Eq}_2)}{A \vdash B_i(\textit{Eq}_1 \wedge \textit{Eq}_2)} \hspace{0.75cm}
 \mathbf{B}\hspace{-0.2em}\triangleright \mprset{sep=1.5em}\inferrule{E \triangleright_i \textit{Eq}_0 \\ \text{for all } \textit{Eq} \in E, A \vdash B_i(\textit{Eq})}{A \vdash B_i(\textit{Eq}_0)}
\end{align*}
\end{mdframed}
\caption{Axiomatics.}
\label{tab:eq:axioms}
\end{table}

The axiomatics meets the following soundness, completeness, and decidability properties.

\begin{prop}[Soundness]
$\text{For all } A \text{ in } \Gamma, \text{ if } A \vdash \phi \text{ then } A \in S \left( \phi \right)$.
\end{prop}

The soundness property can be proved by considering each rule in Table~\ref{tab:eq:axioms} in turn and showing that the traces specified in Table~\ref{tab:eq:semantics-of-architecture} have the expected properties (or that appropriate traces can be found in the case of $\textit{Has}_i^{\textit{all}}$).

\begin{prop}[Completeness]
$\text{For all } A \text{ in } \Gamma, \text{ if } A \in S \left( \phi \right) \text{ then } A \vdash \phi$.
\end{prop} 

Completeness can be proved by systematic inspection of the different cases in Table~\ref{tab:eq:semantics-of-traces-of-events} that can make a property $\phi$ true in the trace semantics.

\begin{prop}[Decidability]
If the deductive systems $\triangleright_i$ are decidable, then the axiomatics is decidable.
\end{prop}

The intuition is that proofs can be stratified into proofs of $\textit{Has}_i^{\textit{all}}$, $\textit{Has}_i^{\textit{none}}$, $\textit{Has}_i^{\textit{one}}$, $\textit{K}_i$, and $\textit{B}_i$ successively, with proofs of properties not involving the deductive systems of the components first and those involving the deductive systems of the components as the last step.

\section{Smart Meter Case Study}
\label{sec:case-study}

One of the services provided by smart metering systems is the periodic billing of an amount $\textit{Fee}$ based on the customers consumption $\textit{Cons}_t$ for periods of time $t$. The service $\textit{Fee} = \sum_t \left( F \left( S \left( \textit{Cons}_t \right) \right) \right)$ (where $F$ and $S$ stand for pricing and metering) is expressed as  $\Omega = \left\{ \textit{Fee} = \odot + (y), y_t = F (x_t), x_t = S ( \textit{Cons}_t ) \right\}$. 
We provide the details for the provider only here but a similar approach could be used for customers or other parties. 

\subsubsection{Architecture Goals.}

The architecture should enable the provider $P$ to get access to the global fee: $A \vdash \textit{Has}_P^{\textit{all}} \left( \textit{Fee} \right)$. However, he should not be able to get access to the individual consumptions $\textit{Cons}_t$ or to the intermediate variables $x$ and $y$ since they are the results of easily inversible functions (typically $F$ is a mapping and $S$ the identity): $A \vdash \textit{Has}_P^{\textit{none}} \left( \textit{Cons} \right) \wedge \textit{Has}_P^{\textit{none}} \left( x \right) \wedge \textit{Has}_P^{\textit{none}} \left( y \right)$. Moreover, he should be convinced that the value provided for $\textit{Fee}$ is actually correct: $A \vdash K_P \left( \textit{Fee} = \odot + (y) \wedge y_t = F (x_t) \wedge x_t = S ( \textit{Cons}_t ) \right)$.

\subsubsection{Architecture Design.}

The design of an architecture meeting the above goals is described Figure~\ref{fig:trust-blind}.
A strong constraint concerning the metering has to be taken into account from the start: regulators generally require the data to be metered by officially certified and tamper-resistant metrological devices $M$: $\textit{Has}_{M} \left( \textit{Cons}\right)$, $\textit{Compute}_{M} \left( x_t = S \left( \textit{Cons}_t \right) \right)$, and $\textit{Attest}_{M} \left( x_t = S \left( \textit{Cons}_t \right) \right)$.

One option for the computation of the fee is to have it performed by the meter: $\textit{Compute}_{M} \left( \textit{Fee} = \odot + (y) \right)$ and $\textit{Compute}_{M} \left( y_t = F(x_t) \right)$. The result of this computation can then be sent to the provider along with the corresponding attestation and the metering attestation through a $\textit{Receive}_{P,M} \left( \left\{ \textit{Att} \right\}, \left\{ \textit{Fee} \right\} \right)$ primitive.
Another architectural primitive $\textit{Verif}^{\textit{Attest}}_P \left( \textit{Att} \right)$ should be added to convince the provider of the correctness of the computation (considering that the provider trusts the meter $\textit{Trust}_{P,M}$).

Finally, the dependance relations have to be defined to model the computational power of the components $P$ and $M$ (they both have the same here for the sake of simplicity, noted $\textit{Dep}_i$ for $i \in \{ P, M \}$). The relations are such that $\left( \textit{Fee}, \left\{ y_t \right\} \right) \in \textit{Dep}_i$, $\left( \textit{y}_t, \left\{ x_t \right\} \right) \in \textit{Dep}_i$, $\left( \textit{x}_t, \left\{ y_t \right\} \right) \in \textit{Dep}_i$, $\left( \textit{x}_t, \left\{ \textit{Cons}_t \right\} \right) \in \textit{Dep}_i$, and $\left( \textit{Cons}_t, \left\{ x_t \right\} \right) \in \textit{Dep}_i$ (only the summation is not inversible here and we have $\left( y_t, \left\{ \textit{Fee} \right\} \right) \notin \textit{Dep}_i$).

\subsubsection{Application of the Axiomatics.}

Rules (H2) and (H6) allow us to prove respectively that the provider gets a value for the global fee since it receives it from the meter and that the consumption and the values of the intermediate variables $x$ and $y$ are not disclosed. Applications of rules (K5) and (K$\wedge$) prove that the correctness of the global fee is ensured thanks to the attestations and the trust relation between the provider and the meter. As expected, (H2) and (H3) prove that the meter has an access to the consumption data.

The solution chosen here for the sake of conciseness describes heavy meters performing the billing computations (which is generally not the case). Moreover, there is a direct link between the meter and the provider: the customer has to trust the meter not to disclose too much data to the provider. This issue could be solved by adding a proxy under the control of the customer which would filter the communications between the provider and the meter. Other options for smart metering such as~\cite{rial:2010} can be expressed in the same framework but space considerations prevent us from presenting them here.

\section{Related Work}
\label{sec:related-works}

This paper stands at the crossroads of three different areas: engineering privacy by design, software architectures and protocols, and epistemic logics.  

Several authors \cite{gurses:2011,kerschbaum:2014,le-metayer:2013,mulligan:2012,spiekermann:2009} have already pointed out the complexity of ``privacy engineering'' as well as the ``richness of the data space''\cite{gurses:2011} calling for the development of more general and systematic methodologies for privacy by design.
As far as privacy mechanisms are concerned, \cite{kerschbaum:2014,maffei:2013} points out the complexity of their implementation and the large number of options that designers have to face. To address this issue and favor the adoption of these tools, \cite{kerschbaum:2014} proposes a number of guidelines for the design of compilers for secure computation and zero-knowledge proofs whereas \cite{fournet:2013} provides a language and a compiler to perform computations on private data by synthesising zero-knowledge protocols. In a different context (designing information systems for the cloud), \cite{manousakis:2013} also proposes implementation techniques to make it easier for developers to take into account privacy and security requirements. 

Software architectures have been an active research topic for several decades \cite{shaw:2006} but they are usually defined using purely graphical, informal means or within semi-formal frameworks. 
Dedicated languages have been proposed to specify privacy properties~\cite{barth:2006,becker:2011,le-metayer:2009,yu:2004} but the policies expressed in these languages are usually more fine-grained than the properties considered here because they are not intended to be used at the architectural level.
Similarly, process calculi such as the applied $\pi$-calculus~\cite{ryan:2011} have been applied to define privacy protocols~\cite{delaune:2010}. Because process calculi are general frameworks to model concurrent systems, they are more powerful than dedicated frameworks.
The downside is that protocols in these languages are expressed at a lower level and the tasks of specifying a protocol and its expected properties are more complex \cite{meadows:2003,paulson:1998,burrows:1990}. Again, the main departure of the approach advocated in this paper with respect to this trend of work is that we reason at the level of architectures, providing ways to express properties without entering into the details of specific protocols that we assume perfect.
The work presented here is a follow-up of~\cite{antignac:2014} which advocates an approach based on formal models of privacy architectures. The framework introduced in~\cite{le-metayer:2013} includes an inference system to reason about the implementation of a ``detectability property'' similar to the integrity property considered here. This framework makes it possible to prove that, in a given architecture, an actor ``A'' can detect potential errors (or frauds) in the computation of a variable ``X''. The logical framework presented here can be seen as a generalisation of~\cite{le-metayer:2013} which does not include a logic for defining privacy and integrity properties.

Epistemic logics have been extensively studied~\cite{fagin:2004}. A difficulty in this kind of framework in a context where hiding functions are used is the problem known as ``logical omniscience''. Several ways to solve this difficulty have been proposed~\cite{pucella:2004,halpern:2007,cohen:2007}. Other works such as \cite{glasgow:1990} also rely on deontic logics and focus on the expression of policies and how they relate to database security or distributed systems.

\section{Directions for Further Work}
\label{sec:directions-for-further-work}

The framework presented in this paper can be used to express in a formal way the main architectural choices in the design of a system and to reason about them. It also makes it possible to compare different options, based on the properties that they comply with, which is of prime importance when privacy requirements have to be reconciled with other, apparently conflicting requirements. 

As stated above, the framework described here does not cover the full development cycle: ongoing work addresses the mapping from the architecture level to the protocol level to ensure that a given implementation, abstracted as an applied $\pi$-calculus protocol~\cite{ryan:2011}, is consistent with an architecture.
Work is also ongoing to integrate this formal framework into a more user-friendly, graphical, design environment integrating a pre-defined design strategy. This strategy, which is implemented as a succession of question-answer iterations, allows the designer to find his way among all possible design options based on key decision factors such as the trust assumptions between entities. The resulting architectures can then be checked using the formal framework described here.

In this paper, we have focused on data minimisation and it should be clear that the framework presented here does not address other privacy requirements such as the purpose limitation or the deletion obligation. Indeed, privacy is a multi-faceted notion that cannot be entirely captured within a single formal framework. Another limitation of the approach is that it must be possible to define the service (or ``purpose'') as the result of a functional expression (e.g. the computation of a fee in electronic toll pricing or smart metering).  Thus the approach does not help in situations such as social networks where the service is just the display of the data (and its access based on a given privacy policy). Last but not least, in this paper, follow a ``logical'' (or qualitative) approach, as opposed to a quantitative approach to privacy and we do not consider the use of auxiliary information. An avenue for further research in this area would be to study the integration of quantitative measures of privacy (such as differential privacy~\cite{dwork:2006a}) into the framework.

\subsubsection{Acknowledgement.}

This work was partially funded by the European project PRIPARE/FP7-ICT-2013-1.5, the ANR project BIOPRIV, and the Inria Project Lab CAPPRIS.

\bibliographystyle{splncs03}
\bibliography{references_shortened}

\end{document}